\documentclass[journal]{IEEEtran}

\usepackage[utf8]{inputenc} 
\usepackage[T1]{fontenc}    
\usepackage{hyperref}       
\usepackage{url}            
\usepackage{booktabs}       
\usepackage{amsfonts}       
\usepackage{nicefrac}       
\usepackage{microtype}      
\usepackage{xcolor}         
\usepackage{graphicx}
\usepackage{subfigure}
\usepackage{amsmath, bm}
\usepackage{makecell}
\usepackage{multirow}
\usepackage{wrapfig}
\usepackage{wrapfig}
\usepackage{fancyvrb}
\usepackage{adjustbox}

\VerbatimFootnotes

\makeatletter

\newcommand{\norm}[1]{\left\lVert#1\right\rVert}

\newcommand{\mean}[1]{\mathbb{E}\left[#1\right]}
\newcommand{\parn}[1]{\left(#1\right)}

\newlength{\NOTskip}

\title{StyleTTS-ZS: Efficient High-Quality Zero-Shot Text-to-Speech Synthesis with Distilled Time-Varying Style Diffusion}

\author{Yinghao~Aaron~Li, Xilin~Jiang, Cong~Han, and~Nima~Mesgarani}

\usepackage{bibentry}

\begin{document}

\maketitle

\begin{abstract}
The rapid development of large-scale text-to-speech (TTS) models has led to significant advancements in modeling diverse speaker prosody and voices. However, these models often face issues such as slow inference speeds, reliance on complex pre-trained neural codec representations, and difficulties in achieving naturalness and high similarity to reference speakers. To address these challenges, this work introduces StyleTTS-ZS, an efficient zero-shot TTS model that leverages distilled time-varying style diffusion to capture diverse speaker identities and prosodies. We propose a novel approach that represents human speech using input text and fixed-length time-varying discrete style codes to capture diverse prosodic variations, trained adversarially with multi-modal discriminators. A diffusion model is then built to sample this time-varying style code for efficient latent diffusion. Using classifier-free guidance, StyleTTS-ZS achieves high similarity to the reference speaker in the style diffusion process. Furthermore, to expedite sampling, the style diffusion model is distilled with perceptual loss using only 10k samples, maintaining speech quality and similarity while reducing inference speed by 90\%. Our model surpasses previous state-of-the-art large-scale zero-shot TTS models in both naturalness and similarity, offering a 10-20× faster sampling speed, making it an attractive alternative for efficient large-scale zero-shot TTS systems. The audio demo, code and models are available at \url{https://styletts-zs.github.io/}. 
\end{abstract}

\section{Introduction}
Recent advancements in text-to-speech (TTS) technology have achieved remarkable progress, bringing TTS systems close to, or even surpassing, human-level performance on various benchmark datasets \cite{tan2024naturalspeech, shen2024naturalspeech, li2024styletts}. With studio-level TTS capabilities nearly perfected, there is a growing demand for more sophisticated tasks such as diverse and personalizable zero-shot speaker adaptation \cite{casanova2022yourtts}. These tasks present a significant challenge due to the need to replicate the unique characteristics and prosodic variations of a vast array of speakers without extensive training data for each individual.

Although there have been rapid developments in zero-shot adaptation, driven by large-scale modeling techniques in large language models (LLMs) \cite{jiang2023mega, wang2023neural, peng2024voicecraft, kim2024clam, chen2024vall}, high-quality discrete audio codecs \cite{zeghidour2021soundstream, defossez2022high, kumar2024high}, and diffusion-based models \cite{shen2024naturalspeech, ju2024naturalspeech, le2024voicebox, lee2024ditto, yang2024simplespeech}, current models face crucial limitations. Many large-scale speech synthesis models rely on auto-regressive modeling \cite{jiang2023mega, wang2023neural, wang2023speechx, jiang2023mega2, peng2024voicecraft, kim2024clam, chen2024vall, meng2024autoregressive}, which results in slower scaling of inference speed as the length of the target speech increases. Alternatively, diffusion-based models are used for building large-scale speech synthesis models \cite{shen2024naturalspeech, le2024voicebox, ju2024naturalspeech, lee2024ditto, yang2024simplespeech, eskimez2024e2}. However, since these models require iterative refinement to produce high-quality results, they also suffer from efficiency issues. Moreover, these models often depend on pre-trained neural codecs not specifically designed for TTS tasks \cite{wang2023neural, wang2023speechx, shen2024naturalspeech}, limiting their ability to naturally model diverse human speech, which encompasses a wide range of speaking styles that can be challenging to control with existing codecs.

In this work, we introduce StyleTTS-ZS, an innovative approach to diverse speech synthesis that aims to address these limitations. Our model decomposes human speech into a global style vector derived from a speaker prompt and prompt-aligned text embeddings that encapsulate the timbre and acoustic features of the speech. Additionally, it includes a fixed-length time-varying style vector that encodes the diverse prosodic variations, such as pitch and duration changes over time. By carefully designing the bottleneck for the style vector space with vector quantization \cite{van2017neural} and multimodal adversarial training \cite{janiczek2024multi}, we can reconstruct speech with high fidelity. We then train a diffusion model \cite{ho2020denoising} to sample the time-varying style vector, effectively modeling the diversity of prosodic variations conditioned on the speaker prompt. This results in efficient latent diffusion \cite{rombach2022high}, as the latent variable is a fixed-length style vector. The simplicity and efficiency of our latent space enable us to distill the teacher diffusion model into a student model using only 10k samples. This distillation maintains diversity and similarity to the prompt speaker while reducing inference to one step.

Our evaluation results demonstrate the effectiveness of StyleTTS-ZS. When trained on the small-scale LibriTTS dataset \cite{zen2019libritts}, our model surpasses several public zero-shot TTS baseline models. Furthermore, when trained on the large-scale LibriLight dataset \cite{kahn2020libri}, comprising 60k hours of data, our model outperforms previous large-scale state-of-the-art (SOTA) TTS models in terms of similarity and naturalness for unseen speakers on the LibriSpeech dataset using only a 3-second reference speaker prompt. Remarkably, StyleTTS-ZS achieves this with nearly 10-20 times faster inference speeds compared to previous SOTA models, showcasing its real-time applicability. 

\section{Related Work}

\begin{figure*}[!ht]
  \centering
  \includegraphics[width=0.9\textwidth]{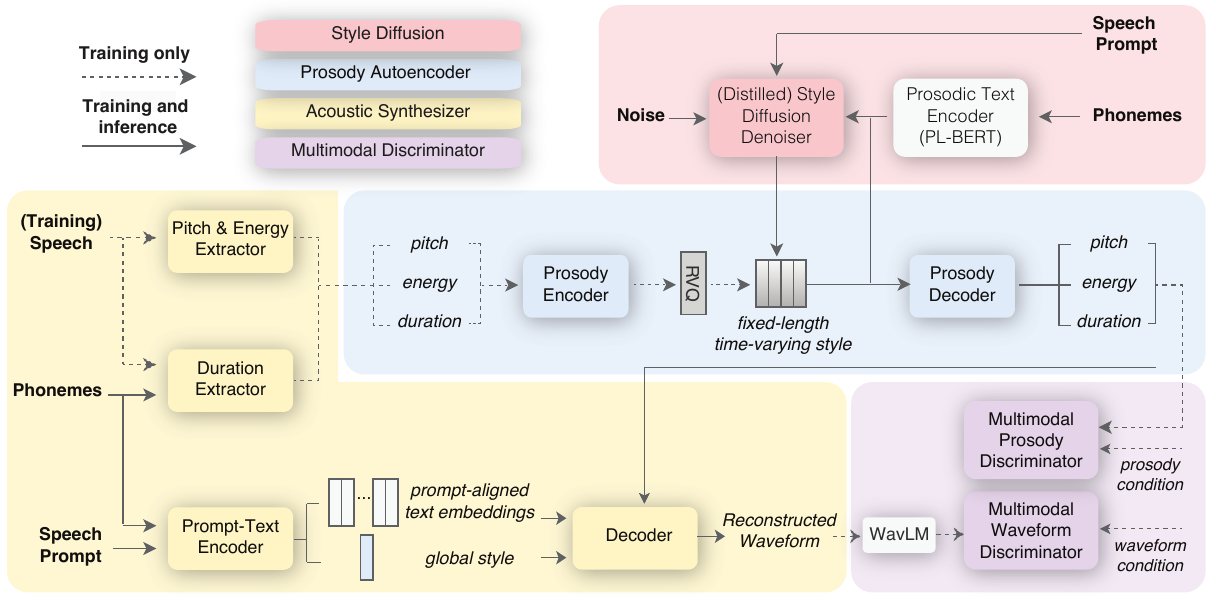}

  \caption{Overview of StyleTTS-ZS architecture. During training, the model uses ground truth speech to extract prosodic features and encode text and style with prompt speech. The prosody encoder compresses these features into a fixed-length time-varying style vector, which is regularized and decoded back by the prosody decoder. The style diffusion denoiser uses this vector for diffusion model training, and the decoder reconstructs speech using prosodic features, text embeddings, and global style, with multimodal discriminators assessing the output. Bold indicates system input, where speech prompts and phonemes are used for both style diffusion and acoustic synthesizer. }
\label{fig:1}

\end{figure*}

\noindent \textbf{Zero-Shot TTS Synthesis. } 
Zero-shot TTS aims to synthesize speech with the voices of unseen speakers using reference speech from the target speaker, making it highly attractive due to its ability to adapt without additional training data. Traditionally, these models are trained on small datasets, using either pre-trained speaker embeddings \cite{casanova2022yourtts, casanova2021sc, wu2022adaspeech, lee2022hierspeech} or speaker encoders in an end-to-end (E2E) manner \cite{li2024styletts, min2021meta, li2022styletts, choi2022nansy++}. However, since the zero-shot TTS task requires generalization to a wide range of unseen speakers, which benefits significantly from large-scale training data, recent advancements in large-scale modeling have shifted focus to in-context learning with reference prompts \cite{wang2023neural}, employing either auto-regressive models, such as large language models to predict subsequent speech tokens \cite{shen2024naturalspeech, le2024voicebox, ju2024naturalspeech, lee2024ditto, yang2024simplespeech, eskimez2024e2}, or non-autoregressive approaches that predict the whole speech given part of speech as a prompt, which often involve diffusion-based techniques for iterative refinement to achieve high-quality results \cite{jiang2023mega, wang2023neural, wang2023speechx, jiang2023mega2, peng2024voicecraft, kim2024clam, chen2024vall, meng2024autoregressive, yang2024simplespeech, lovelace2023simple}. Our approach combines traditional encoder methods and in-context learning by modeling speech as prompt-aligned text embeddings to capture local acoustic features with a joint text-prompt encoder while predicting the more global time-varying style vector of the whole speech with diffusion-based models using prompt speech for diverse prosodic variations.

\noindent\textbf{Efficient High-Quality Speech Synthesis. }  Autoregressive models have been shown to produce expressive and diverse speech \cite{wang2023neural}, but they are prone to erroneous enunciations \cite{song2024ella} and suffer from slow inference speeds proportional to the duration of the target speech. Non-autoregressive approaches \cite{ren2020fastspeech} accelerate inference by generating speech in parallel, but often lack fine details, resulting in blurred speech or flat prosody. These issues have been addressed by adversarial training \cite{kim2021conditional} and diffusion models \cite{popov2021grad}, although diffusion models introduce additional inference time. Recent efforts to speed up diffusion-based models include distilling pre-trained diffusion models \cite{huang2022prodiff, ye2023comospeech} or using consistency \cite{ye2024flashspeech} or rectified flow matching training \cite{guan2024reflow}. However, since these models operate in the design space of mel-spectrograms or pre-trained neural codecs not specifically crafted for TTS synthesis, there is a trade-off between quality and inference steps due to the "no free lunch principle" \cite{ho2002simple}, where attempts to speed up models generally result in a loss of quality or diversity, as seen in recent diffusion distillation works \cite{luo2023comprehensive}. Motivated by the fact that acoustic features are stable and do not need to be modeled stochastically using a diffusion model, while prosody varies over time and hence benefits from diffusion modeling, StyleTTS-ZS carefully designs the latent space only to include prosody information to lessen the burden of diffusion models and minimizes quality loss during distillation. By balancing reconstruction error of autoencoder and diffusion model complexity, we identify a simple and structured latent space for diffusion and then distill the diffusion model in a single inference step. As a result, StyleTTS-ZS significantly outperforms a very recent large-scale efficient high-quality TTS model \cite{ye2024flashspeech} while maintaining similar inference speed.

\section{StyleTTS-ZS}
StyleTTS-ZS consists of four modules: acoustic synthesizer, prosody autoencoder, time-varying style diffusion, and multimodal discriminators. We detail these four modules in the following sections with an overview of our framework in Figure \ref{fig:1} and implementation details in Appendix \ref{app:C} and \ref{app:D}. 

\subsection{Acoustic Synthesizer}
\label{sec:3.1}
The role of the acoustic synthesizer is to reconstruct input speech \(\bm{x}\) using its text transcription \(\bm{t}\) and a speech prompt \(\bm{x}'\) from the same speaker into the reconstructed speech \(\hat{\bm{x}}\). This process starts with extracting pitch \(p\), energy \(n\), and duration \(d\) from the input speech \(\bm{x}\). The joint prompt-text encoder \(T\) then encodes the phoneme text \(\bm{t}\) and the prompt speech \(\bm{x}'\) into prompt-aligned text embeddings \(\bm{h}_{\text{text}} = T(\bm{t}, \bm{x}')\) and a global style \(\bm{s}\). The speech is then reconstructed as \(\hat{\bm{x}} = G(\bm{h}_{\text{text}}, p, n, d, \bm{s})\).

We use the same pitch extractor $F$, duration extractor $A$, and decoder $G$ as in \cite{li2024styletts}. Unlike previous works that rely solely on global speaker embeddings or style vectors \cite{min2021meta, casanova2022yourtts, li2022styletts, li2024styletts} or prompt-aligned text embeddings \cite{huang2022generspeech, kim2024p}, we combine both approaches and jointly encode the speech prompt and text to produce both prompt-aligned text embeddings and a style vector ({Figure \ref{fig:2}c}). Our prompt-text encoder is similar to \cite{kim2024p} but uses conformer blocks \cite{gulati2020conformer} instead of transformer blocks \cite{vaswani2017attention} to better model the speech. Moreover, instead of discarding the output portion from the speech input, we apply average pooling and use the pooled results as the global style. Utilizing both prompt-aligned text embeddings and global style vectors enhances speaker similarity, as demonstrated in Section { \ref{sec:ab}}. 

When training on large-scale datasets with thousands of speakers, we observed that the mel-spectrogram reconstruction loss alone was insufficient for achieving high-fidelity voice similarity due to the limited capacity of the acoustic synthesizer, which does not use transformers to optimize the inference speed. To address this, we introduced an additional reconstruction loss that aligns with the intermediate features of speaker embedding models \cite{wang2023wespeaker}. Since our model operates directly in the waveform domain and generates waveforms end-to-end without relying on any codec decoder or vocoder, we can compute and match these intermediate features directly within the speaker embedding space. This significantly enhances speaker similarity, as shown in Section {\ref{sec:ab}}.

The AdaIN-based decoder from \cite{li2024styletts} is easier to train than the attention-based module in the prompt-text encoder, causing the model to over-rely on the global style vector. Consequently, the prompt-aligned text embeddings become too similar to plain text embeddings. To mitigate this, we apply a 20\% dropout to the global style vector during training, which compels the prompt-text encoder to focus more on aligning text with the prompt and producing a global style. This strategy enhances reconstruction quality by ensuring the prompt-text encoder actively contributes to acoustic synthesis.

\subsection{Vector Quantized Prosody Autoencoder}
While the acoustic synthesizer can reconstruct speech from prompt-aligned text embeddings, phoneme duration, pitch, energy, and a global style vector with high fidelity, we lack ground truth for these prosodic features during inference. Predicting these features from text alone is challenging due to their variability and diversity, especially in large-scale datasets with various speakers. Prosodic features are crucial for both speech naturalness and speaker similarity; unnatural prosody can make speech sound robotic, while uncharacteristic prosody produces dissimilar voices to the prompt speaker despite having the same timbre. Recent works in large-scale TTS models attempt to model this variability on large datasets using methods from large language models (LLMs) \cite{jiang2023mega, jiang2023mega2} and large diffusion models \cite{ju2024naturalspeech}. However, these methods are inefficient as the computation required to infer prosody is proportional to the length of the target speech. We take an innovative approach to compressing these prosodic features into a fixed-length time-varying style vector via a prosody autoencoder, allowing efficient diffusion sampling for diverse prosody of human speech.

\begin{figure*}[!t]

\vspace{0 pt}
  \centering
  \includegraphics[width=0.9\textwidth]{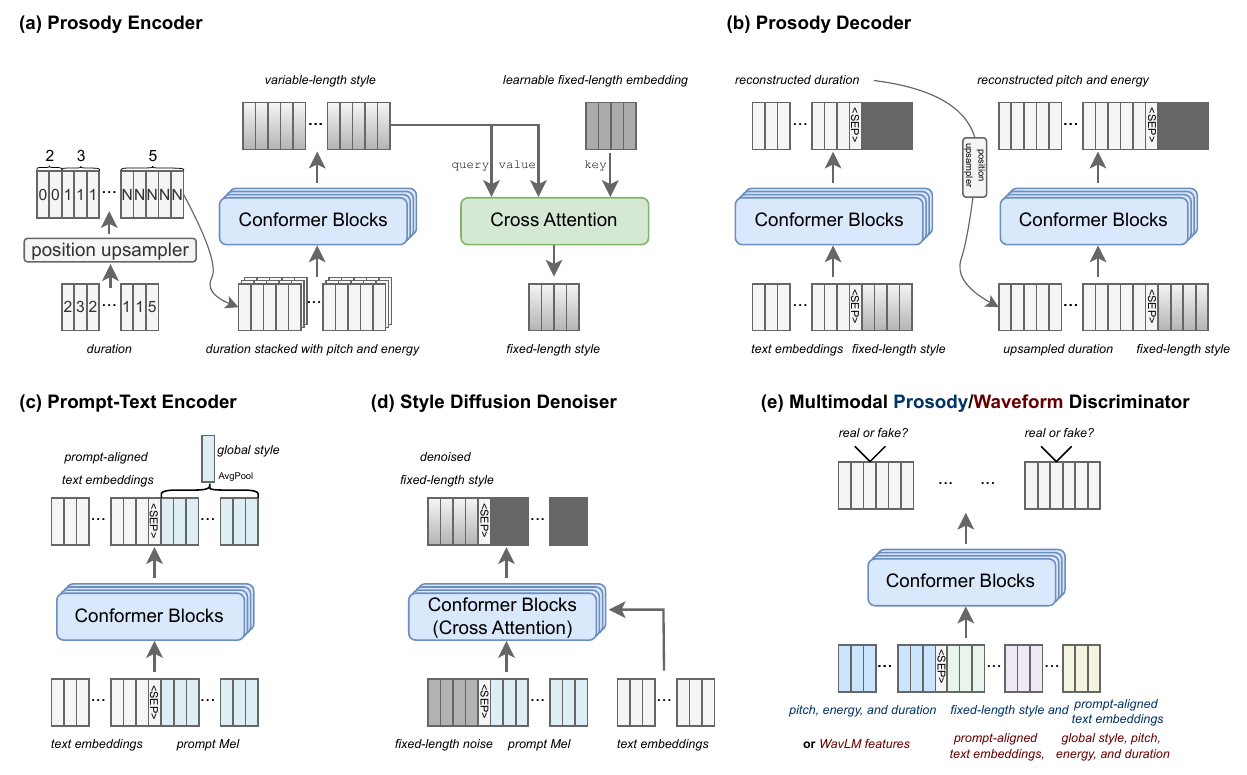}

  \caption{Architectures for newly proposed components in StyleTTS-ZS. For (b) and (d), the dark part of the output means this part is discarded for output, and only the grey part is used.  }
\label{fig:2}

\end{figure*}

The prosody encoder ({Figure \ref{fig:2}a}) processes stacked pitch, energy, and duration inputs, mapping them into a fixed-length vector. We represent duration using upsampled positional embeddings \(\text{PE}(\cdot)\), where for each \(i \in \{1, \ldots, N\}\), the positional embedding \(\text{PE}(i)\) is repeated \(d_i\) times. This representation efficiently distinguishes each phoneme's duration without complicating the latent space by taking additional text embeddings as input. The stacked prosody representations are fed into conformer blocks to extract variable-length prosody representations \(\bm{h}_{\text{vl}}\). To compress these into a fixed-length vector, we use cross-attention \(\text{Attn}(k, q, v)\) with \(\bm{h}_{\text{vl}}\) as the query and value, and learnable fixed-length positional embeddings \(\bm{h}_{\text{pe}}\) as the key. This yields the final latent \(\bm{h}_{\text{style}} = \text{Attn}(\bm{h}_{\text{pe}}, \bm{h}_{\text{vl}}, \bm{h}_{\text{vl}})\), which we name \textit{time-varying style} to differentiate it from previous works that use global style vectors to represent speech styles. The prosody decoder $\text{PD}(\cdot)$ ({Figure \ref{fig:2}b}) then uses \(\bm{h}_{\text{style}}\) to decode the duration $\hat{d}$, pitch $\hat{p}$, and energy $\hat{n}$ conditioned on the PL-BERT \cite{li2023phoneme} phoneme embeddings from \(\bm{t}\). 

We noticed that this method achieves almost perfect prosody reconstruction with minimal perceptual difference from the input, even with a vector length of \(K = 50\) for up to 30 seconds of speech without adversarial training. However, this leads to a latent space that is overly detailed, making diffusion modeling and one-step distillation difficult. To address this, we apply residual vector quantization (RVQ) \cite{zeghidour2021soundstream}, simplifying the latent space by quantizing it. This reduces details in the latent space and simplifies the diffusion model's task at the cost of reconstruction accuracy of the autoencoder, which can be mitigated using adversarial training with multimodal discriminators. We use 9 codebooks with 1024 codes each and project the time-varying style with \(d=512\) into a lower space with \(d=8\) for efficient quantization following \cite{kumar2024high}, achieving a balance between diffusion model difficulty and prosody reconstruction fidelity (see Appendix  \ref{app:a2} for more discussions).

To further shift the burden of the diffusion model to the prosody decoder, we randomly mask the input pitch, energy, and duration fed to the prosody encoder. This technique encourages the prosody decoder to learn to reconstruct prosodic features from partial input, which is particularly beneficial for zero-shot TTS scenarios where the input prompt is short. 

\subsection{Distilled Time-Varying Style Diffusion}
\label{sec:3.3}
We sample the latent $\bm{h}_\text{style}$ which we denote as $\bm{h}$ for the rest of this section conditioned on the PL-BERT embeddings from input text $\bm{t}$ and speech prompt $\bm{x}'$ by modeling $p(\bm{h}|\bm{t}, \bm{x}')$ using a diffusion model. We follow the probability flow ODE formulation in \cite{song2020score} for deterministic sampling: 

\begin{equation}
\begin{split}
\label{eq:1}
    d\bm{h} = & \left[f(\bm{h}, \tau) - \frac{1}{2}g^2(\tau) \nabla_{\bm{h}} \log p_\tau(\bm{h}|\bm{t}, \bm{x}') \right]\,d\tau, \\
    & \bm{h}(1)  \sim \mathcal{N}(0, \sigma_1^2I),
    \end{split}
\end{equation}
where $f(\bm{h}, \tau) := \frac{d}{d\tau} \log \alpha_{\tau}$ is the drift coefficient, $g(\cdot, \tau) := \frac{d}{d\tau}\sigma_\tau^2 \parn{1 - 2 \log\parn{\alpha_{\tau}}}$ is the diffusion coefficient, $\sigma_\tau$ is the noise level, $\alpha_{\tau}$ is the schedule for $\sigma_\tau$, and $\nabla_{\bm{h}} \log p_\tau(\bm{h}|\bm{t}, \bm{x}')$ is the score function at time $\tau$, estimated using a denoiser $K(\cdot\; ; \sigma_\tau, \bm{t}, \bm{x}')$ with architecture in {Figure \ref{fig:2}d}: 

\begin{equation}
\label{eq:2}
    \nabla_{\bm{h}} \log p_\tau(\bm{h}|\bm{t}, \bm{x}') = \frac{\alpha_{\tau}K(\bm{h}(\tau); \sigma_\tau, \bm{t}, \bm{x}') - \bm{h}(\tau)}{{\sigma_\tau}}.
\end{equation}
We train the denoiser using the velocity formulation \cite{salimans2022progressive} with the following objective:
\begin{equation}
\begin{split}
\label{eq:3}
    \mathcal{L}_\text{diff} = \mathbb{E}_{\bm{x}, \bm{x'}, \bm{t}, \tau \sim \mathcal{U}([0, 1]), \bm{\xi} \sim \mathcal{N}(0, I)}[\lVert K(\alpha_{\tau}E(\bm{x}) + \\ \sigma_\tau\bm{\xi}; \theta, \sigma_\tau, \bm{t}, \bm{x}') - \bm{v}(\sigma_\tau, E(\bm{x})) \rVert_1],
\end{split}
\end{equation}
where $E(\cdot): \mathcal{X} \rightarrow \mathcal{H}$ denotes combined pitch, energy and duration extractor and prosody encoder that maps speech $\bm{x} \in \mathcal{X}$ to latent $\bm{h} \in \mathcal{H}$. The velocity $\bm{v}$ is defined as $\bm{v}(\sigma_\tau, x) := \alpha_{\tau} \bm{\xi} - \sigma_\tau x$, with an angular scheduler $\alpha_{\tau} := \cos(\phi_\tau)$ and $\sigma_\tau := \sin(\phi_\tau)$ for $\phi_\tau = \frac{\pi}{2}\tau$. 
\begin{table*}[!th]
\small

  \caption[Caption for LOF] {Comparative mean opinion scores of naturalness (CMOS-N) and similarity (COMS-S) for StyleTTS-ZS (LL) relative to other models (positive scores indicate StyleTTS-SZ is better; one asterisk * indicates $p < 0.05$ and two asterisks ** indicate $p < 0.01$), predicted MOS (UT-MOS), speaker embedding similarity (SIM), word error rate (WER), coefficient of variation for pitch and energy ($\text{CV}_{p + n}$) and real-time factor (RTF)\textsuperscript{\protect\ref{secondfootnote}} in comparison to other recent large-scale models and StyleTTS-ZS (LT).}

  \label{tab:1}
  \centering
  \begin{tabular}{lccccccccc}
    \toprule
    Model & Training Set & CMOS-N  & CMOS-S & UT-MOS $\uparrow$ & WER $\downarrow$ & SIM $\uparrow$ &$\text{CV}_{p + n}$ $\uparrow$ & RTF $\downarrow$   \\ 
    \midrule
    Ground Truth  & ---   & $-0.44^{*}$ & $0.77^{**}$ & 4.17 & 0.34 & 0.67 & 1.49 & ---  \\
    \midrule
    Vall-E \cite{wang2023neural}  & LibriLight   &  $1.07^{**}$ & $0.65^{**}$  & 3.31 & 4.97 & 0.47 &  0.94 & 0.62 \textsuperscript{\textdagger} \\
    NaturalSpeech 2 \cite{shen2024naturalspeech} & MLS   & $0.57^{**}$  & $0.19^{*}$ & 3.78 & 1.25  & {0.55} & {1.07} & 0.37 \textsuperscript{$\ddagger$}  \\
    
    
    VoiceCraft \cite{peng2024voicecraft} &  GigaSpeech  & $0.84^{**}$  & $0.11^{*}$ & 3.58 & 3.73 & {0.54} &  0.79 & 1.24 \textsuperscript{$\ddagger$}  \\
    FlashSpeech \cite{ye2024flashspeech} &  MLS  & $0.42^{*}$  &  $0.52^{**}$ & 3.98 &  1.47  & {0.50} &  0.50 & \textbf{0.02} \textsuperscript{$\ddagger$} \\ 
    NaturalSpeech 3 \cite{ju2024naturalspeech} &  LibriLight  & $0.28^{*}$  &  0.01 & 4.09 & 1.81  & \textbf {0.66} &  1.23 & 0.30
    \textsuperscript{$\ddagger$} \\
    \midrule
    StyleTTS-ZS (LT)  & LibriTTS   &  $0.21^{*}$  & $0.31^{*}$ & \textbf{4.24} &  0.90 & 0.47 & 1.06 & {0.03}  \textsuperscript{$\ddagger$} \\
    StyleTTS-ZS (LL) & LibriLight   &  \textbf{0.00}  & \textbf{0.00} & {4.16} & \textbf{0.79} &  0.56 &  \textbf{1.67} & {0.03}  \textsuperscript{$\ddagger$} \\

    \bottomrule
    
  \end{tabular}
  
\end{table*}
We apply classifier-free guidance (CFG) \cite{ho2022classifier} using both speech prompt $\bm{x}'$ and text $\bm{t}$ as a condition. The modified denoiser with CFG is:
\begin{equation}
\begin{split}
\tilde{K} & (\cdot \;; \omega, \sigma_{n}, \bm{t}, \bm{x}')  := K(\cdot \;; \sigma_{n},  \emptyset)  + \\ & \omega \cdot \parn{K(\cdot \;;  \sigma_{n}, \bm{t}, \bm{x}') - K(\cdot \;;  \sigma_{n}, \emptyset)},
\end{split}
\end{equation}
where \(\omega\) is the guidance scale and \(\emptyset\) indicates null condition embeddings. We randomly dropped out the condition $\bm{x}'$ with rate of 0.1 during training and fixed $\omega = 5$ during inference. 

Equation $\ref{eq:1}$ can be viewed as a neural ODE following a trajectory that maps a Gaussian noise $\xi \in \mathcal{N}(0, I)$ to the time-varying style $\bm{h} \in \mathcal{H}$ conditioned on $\bm{x}'$ and $\bm{t}$. We denote this map as $h(\cdot\; ;\omega, \bm{x}', \bm{t}): \mathcal{N} \rightarrow \mathcal{H}$. We solve this ODE using the deterministic solver DDIM \cite{song2020denoising} for later distillation that repeatedly applies the followings for $n \in \{L, L - 1, \ldots, 0\}$:
\begin{equation}
\label{eq:4}
\begin{split}
    \bm{h}_{\sigma_{n - 1}} = \alpha_{n - 1}{\bm{v}}_n + \sigma_{n-1} \tilde{\bm{v}}_{n},
\end{split}
\end{equation}
where ${\bm{v}}_n := \alpha_{n}{\bm{v}}_n - \sigma_{n} \tilde{K}(\bm{h}_{\sigma_{n}}; \omega, \sigma_{n}, \bm{t}, \bm{x}')$ and $\tilde{\bm{v}}_n = \sigma_{n}{\bm{v}}_n + \alpha_{n} \tilde{K}(\bm{h}_{\sigma_{n}}; \omega, \sigma_{n}, \bm{t}, \bm{x}')$ for $L = 100$. 

We can train a student network $H(\cdot\; ; \omega, \bm{x}', \bm{t})$ to approximate $h$ as it is a deterministic map. Since obtaining samples to train $H$ through equation \ref{eq:4} can be expensive especially with large integration steps $L = 100$ due to the need to accurately reflect the effects of CFG, we initialize our student network using a pre-trained network $H'$ that predicts prosody decoder output from the text $\bm{t}$ and a speech prompt $\bm{x}'$. This student initialization allows sample reduction to as small as 10k samples as shown in Appendix \ref{app:a3}, since the model has already learned a deterministic map from the speech prompt and text to the time-varying style as we match the initial student network and target distilled diffusion sampler in the output space. The student network can be pre-trained during the style diffusion training. We use perpetual loss for distillation \cite{liu2023instaflow}, for which our perpetual metric is the prosody decoder's output. The distillation loss is defined as:

\begin{equation}
\label{eq:5}
\begin{split}
    \mathcal{L}_\text{dist} = \mathbb{E}_{\substack{\bm{x'}, \bm{t}, \bm{\xi} \sim \mathcal{N}(0, I), \\ \omega \in \mathcal{U}([1, 15])}} & [\lvert\text{PD}\parn{H(\bm{\xi}; \omega, \bm{x}', \bm{t})} \\ & - \text{PD}\parn{h(\bm{\xi}; \omega, \bm{x}', \bm{t})}\rvert_1].
\end{split}
\end{equation}
We show this simulation-based approach for distillation is superior to other simulation-free methods that use bootstrapping, such as consistency distillation \cite{song2023consistency} and adversarial diffusion distillation \cite{sauer2023adversarial} in Appendix \ref{app:a3}. This is because our latent space is a $50 \times 512$ vector and can be sampled fairly fast and distilled with 10k samples, which took a few hours to obtain on 8 NVIDIA RTX 3090 GPUs.

\subsection{Multimodal Discriminators}
\label{sec:3.4}
One observation we made is the trade-off between latent space complexity, reconstruction error, and the difficulty of diffusion model training and distillation (see Appendix \ref{app:a2}). A simpler latent space makes the diffusion model easier to train and distill but increases the reconstruction error for the autoencoder. To achieve efficient latent diffusion and high-fidelity distillation, we opted to simplify the latent space for the time-varying style and optimize the autoencoder for high-quality reconstruction. We introduce multimodal discriminators that evaluate not only the decoder output to determine if the sample is real or fake but also consider the decoder input as an additional modality, which has been shown to improve speech quality for zero-shot speech synthesis \cite{janiczek2024multi}.

Specifically, we use two multimodal discriminators ({Figure \ref{fig:2}e}): one for the waveform decoder and one for the prosody decoder. The waveform discriminator takes the WavLM \cite{chen2022wavlm} features of the decoder's output following the idea of SLM discriminator in \cite{li2024styletts} that has been subsequently demonstrated effectively \cite{li2023slmgan, ye2024flashspeech} and conditions on all the inputs to the decoder (prompt-aligned text embeddings, global style, pitch, energy, and duration). The prosody discriminator takes the prosody decoder's output and conditions on all the inputs to the prosody decoder, including the time-varying style and PL-BERT text embeddings. This approach significantly enhances the naturalness and similarity of the reconstructed speech, as demonstrated in {Section \ref{sec:ab}}. In addition to the multimodal discriminator, we also have the multi-period discriminator (MPD) \cite{kong2020hifi} and multi-resolution STFT discriminator (MRD) \cite{kumar2024high} for the waveform decoder.

\section{Experiments}
\label{sec:4}
\subsection{Model Training}
We performed experiments on two datasets: LibriTTS and LibriLight. First, we trained our model on the small-scale LibriTTS dataset \cite{zen2019libritts}, which contains about 585 hours of audio from 1,185 speakers. Utterances longer than 30 seconds or shorter than one second were excluded. The training set comprised the combined \textit{train-clean-100}, \textit{train-clean-360}, and \textit{train-other-500} subsets. Next, we trained our model on the large-scale LibriLight dataset \cite{kahn2020libri}, consisting of 57,706.4 hours of audio from 7,439 speakers. Since this dataset lacks text transcriptions, we obtained them using methods from \cite{kang2024libriheavy}. All datasets were resampled to 24 kHz to match LibriTTS, and texts were converted into phonemes using Phonemizer \cite{Bernard2021}. We randomly truncated the input speech to the smallest length in the batch and used the truncated speech as prompts during training. 

We trained our model for 30 epochs on the LibriTTS dataset and 1 million steps on the LibriLight dataset. After training, we distilled the style diffusion denoiser using 10k samples with texts and prompts randomly sampled from the training set, and trained the student model for 10 epochs. We employed the AdamW optimizer \cite{loshchilov2018fixing} with \(\beta_1 = 0\), \(\beta_2 = 0.99\), weight decay \(\lambda = 10^{-4}\), learning rate \(\gamma = 10^{-4}\), and a batch size of 32 samples. Since waveforms require more GPU RAM than prosody features, for acoustic synthesizer training, waveforms were randomly segmented to a maximum length of 3 seconds, while the prosody autoencoder and style diffusion denoiser were trained with full audio lengths up to 30 seconds. Training was conducted on four NVIDIA L40 GPUs.

\subsection{Evaluations}
\label{sec:4.2}
We employed two metrics in our experiments: Mean Opinion Score of Naturalness (MOS-N) for human-likeness, and Mean Opinion Score of Similarity (MOS-S) for similarity to the prompt speaker. These evaluations were conducted by native English speakers from the U.S. on Amazon Mechanical Turk. All evaluators reported normal hearing and provided informed consent. 
We conducted two experiments with different groups of baseline models: one for small-scale models and another for large-scale models.

For small-scale models, we compared our model to three high-performing public models: XTTS-v2 \cite{casanova2022yourtts}, StyleTTS 2 \cite{li2024styletts}, and HierSpeech++ \cite{lee2023hierspeech++} on the LibriTTS dataset. Each synthesized speech set was rated by 10 evaluators on a 1-5 scale, with 0.5 increments. We randomized the model order and kept their labels hidden, similar to the MUSHRA approach \cite{li2021starganv2, li2022styletts}. We tested 40 samples from the LibriSpeech \cite{panayotov2015librispeech} \textit{test-clean} subset with 3-second refernece speech, following \cite{wang2023neural}. Official checkpoints trained on LibriTTS were used for all baseline models (see Appendix \ref{app:B1} for more information about baseline models).

\begin{table}[!t]
\setlength{\tabcolsep}{1mm} 
\small
  \caption{Comparison of MOS with 95\% confidence intervals (CI), word error rate (WER) and real-time factor (RTF) for public models trained on LibriTTS. StyleTTS-ZS uses LT model, and StyleTTS 2 uses 5 steps for style diffusion. }
  \label{tab:2}
  \centering

  \begin{tabular}{lllcc}
    \toprule
    Model  & MOS-N (CI) $\uparrow$ & MOS-S (CI) $\uparrow$ & WER $\downarrow$ & RTF $\downarrow$ \\ 
    \midrule
    Ground Truth              & 4.67 ($\pm$ 0.11) & 4.32 ($\pm$ 0.10) & 0.34 & --- \\
    StyleTTS-ZS     & \textbf{4.54} ($\pm$ \textbf{0.11}) & \textbf{4.33} ($\pm$ \textbf{0.11}) & \textbf{0.90} & \textbf{0.0320} \\
    StyleTTS 2 & 4.23 ($\pm$ 0.11)  & 3.42 ($\pm$ 0.09) & 1.61 & {0.0671}\\
    HierSpeech++            & 3.54 ($\pm$ 0.12) & 4.27 ($\pm$ 0.12) & 7.82 & 0.1969\\
    XTTSv2      &  3.68 ($\pm$ 0.09) & 3.74 ($\pm$ 0.10) & 6.17 & 0.3861 \\
    \bottomrule
  \end{tabular}
\end{table} 

For large-scale experiments, since most state-of-the-art large-scale models are not publicly available, we compared our model to audio samples obtained from the authors or official demo pages using comparative MOS (CMOS) tests, as raters can ignore subtle differences in MOS experiments, making it difficult to estimate accurate performance from limited samples. Raters compared pairs of samples and rated whether the second was better or worse (or more or less similar to the prompt speaker) than the first on a -6 to 6 scale, with 1-point increments. We included five recent models: Vall-E, NaturalSpeech 2, NaturalSpeech 3, FlashSpeech, and VoiceCraft. For Vall-E, NaturalSpeech 2/3 and FlashSpeech, we obtained 40 samples from the authors with 3-second of prompt speech of unseen speakers in LibriSpeech \textit{test-clean} subset and used these samples for evaluations.  Since the model of VoiceCraft is publicly available, we synthesized the samples using the same 40 prompts and texts. 

In addition to subjective evaluations, we followed \cite{shen2024naturalspeech, ju2024naturalspeech, ye2024flashspeech} for objective evaluations of sound quality using predicted MOS (UT-MOS) \cite{saeki2022utmos},  robustness using word error rate (WER) from a pre-trained ASR model \footnote{\url{https://huggingface.co/facebook/hubert-large-ls960-ft}} and similarity to the reference speaker (SIM) by cosine similarity from a pre-trained speaker verification model \footnote{\url{https://github.com/microsoft/UniSpeech/tree/main/downstreams/speaker_verification}}. We also measured the pitch and energy coefficient of variation following \cite{li2024styletts} for expressiveness. Additionally, we measured prosody similarity by computing the Pearson correlation coefficients of acoustic features associated with emotions and speech duration between the prompt and the synthesized speech, following \cite{li2022styletts}. 

\section{Results}

\begin{table*}[!ht]
\small
\centering
\caption{Ablation study on LibriTTS for verifying the effectiveness of each proposed component. Significant results ($p < 0.05$) are marked by an asterisk (*). For w/o distillation, the RTF is 0.28. }
\setlength{\tabcolsep}{1mm} 

\label{tab:4}
    \begin{adjustbox}{width=\columnwidth}
    \begin{subtable}
    \centering

    \begin{tabular}{lcccccr}
    \toprule
    Model & CMOS-N & CMOS-S & UT-MOS & SIM & WER \\ 
    \midrule
    StyleTTS-ZS (LT)                 & {0}  & {0} & 4.24 & 0.47 & {0.90}  \\
    \midrule
    w/o PATE           & $-0.24^{*}$ & $-0.18^{*}$ & 3.98 & 0.38 & 1.14 \\
    w/o global style           & $-0.31^{*}$ & $-0.47^{*}$  & 3.54 & 0.34 & 1.45 \\
    w/o SEFM Loss           & $-0.02$ & $-0.23^{*}$ & 4.31 & 0.41 & 0.10  \\
    \bottomrule
    \end{tabular}    
    \end{subtable}
    \end{adjustbox}
    \begin{adjustbox}{width=\columnwidth}
\hfil
    \begin{subtable}
    \centering

    \begin{tabular}{lcccccr}
    \toprule
    Model & CMOS-N & CMOS-S & UT-MOS & SIM & WER \\ 
    \midrule
    StyleTTS-ZS  (LT)                 & {0}  & {0} & 4.24 & 0.47 & {0.90}  \\
    \midrule
    w/o distillation           & $-0.02$ & $0.06$ & 4.12 & 0.43 & 0.90 \\
    w/o MMWD           & $-0.24^{*}$ & $-0.29^{*}$ & 3.97 & 0.39 & 1.22 \\
    w/o MMPD           & $-0.58^{*}$ & $-0.32^{*}$ & 4.19 & 0.40 & 0.96 \\
    \bottomrule
    \end{tabular}
    \end{subtable}
    \end{adjustbox}
\label{tab:ab}

\end{table*}

\subsection{Model Performance}
\label{sec:5.1}

\footnotetext[3]{\label{secondfootnote} \textdagger: device unknown for RTF and results are taken from the original paper. $\ddagger$: RTF was computed on an NVIDIA V100 GPU.}

As shown in Table \ref{tab:1}, our model trained on the large dataset has outperformed previous state-of-the-art (SOTA) large-scale TTS models in multiple metrics: human rated naturalness (CMOS-N), predicted sound quality (UT-MOS), similarity (CMOS-S), expressiveness (CV), inference time (RTF), and robustness (as indicated by WER). We note that we achieve competitive performance in SIM with most models except NaturalSpeech 3 despite having a statistically insignificant CMOS-S compared to it. This may be due to our model's adversarial training with multimodal discriminators, which enhances speaker likeness from a human perception perspective, whereas other models use a pre-trained neural codec that aligns more with neural network perceptions but not necessarily human perceptions \cite{ju2024naturalspeech}. Although the current SOTA NaturalSpeech 3 has achieved ground-truth level performance in terms of similarity measured by speaker verification models, it still falls short of robustness and naturalness where our model excels. Notably, our model has demonstrated similar performance in terms of human-perceived similarity as NaturalSpeech 3 and has achieved similarly superior perceived similarity than ground truth. In addition, our model is 10$\times$ faster than NaturalSpeech 3.

Since our model does not use iterative refinement methods, it is among the fastest large-scale TTS models, second only to FlashSpeech in efficiency, while significantly surpassing it in both naturalness and similarity. Additionally, our model exhibits greater robustness than all other models, as indicated by the WER scores. The expressiveness of our model, shown by the pitch and energy standard deviation, indicates that it closely matches the ground truth in terms of speech variation and expressiveness. We also include an breakdown of time taken for each module in Appendix \ref{app:a5}.

Our model also outperforms other public models on small-scale data with only 585 hours of audio, as shown in Table \ref{tab:2}. Moreover, when comparing our model trained on larger data (StyleTTS-ZS LL), both CMOS-N and CMOS-S scores are significantly higher than the model trained on smaller data (StyleTTS-ZS LT), despite using the same amount of parameters. This demonstrates our model’s scalability and capability to handle larger datasets effectively. In Table \ref{tab:3}, we see that our model has outperformed other zero-shot TTS models in most of the acoustic characteristics associated with emotions, demonstrating its ability to reproduce the speech style of prompt speech.  

\subsection{Ablation Study} 
\label{sec:ab}

To verify the effectiveness of each proposed component, we conducted extensive ablation studies on LibriTTS using two subjective metrics, CMOS-N for naturalness and CMOS-S for similarity, and evaluated UT-MOS, speaker embedding similarity (SIM) and word error rate (WER) for robustness. We used the same 40 texts and prompt speech as in the other MOS experiments and tested the following variations:

\begin{itemize}
\item \textit{w/o PATE}: Using only global styles in acoustic synthesizer without prompt-aligned text embeddings (PATE). In this case, global styles are computed along with the prompt, but the text embeddings are computed with a prompt of value all 0.
\item \textit{w/o global style}: Using only prompt-aligned embeddings in the acoustic synthesizer without global styles. All AdaIN layers in the decoder were replaced with instance normalization. 
\item \textit{w/o SEFM Loss}: No speaker embedding feature matching (SEFM) loss as in eq. \ref{eq:sv}. 
\item \textit{w/o distillation}: Using the original diffusion model instead of distilled one for inference.
\item \textit{w/o MMWD}: No multimodal waveform discriminator (MMWD) for acoustic synthesizer.
\item \textit{w/o MMPD}: No multimodal prosody discriminator (MMPD) for the prosody autoencoder.
\end{itemize}

As shown in Table \ref{tab:ab}, both prompt-aligned text embeddings and global style vectors are crucial for high-quality speech synthesis with high fidelity to the prompt speaker, with global style being more important likely because the AdaIN-based decoder benefits significantly from the global style, as demonstrated in \cite{li2022styletts}. Omitting classifier-free guidance results in decreased speaker similarity, as the prosody becomes less characteristic of the prompt, though it slightly lowers the word error rate. This is likely because speaker-characteristic duration causes the ASR model to misrecognize certain words, but it has minimal impact on perceived naturalness. Using the distilled instead of the original diffusion model has minimal impact on perceived naturalness and similarity with even a boost of predicted MOS due to mode shrinkage during distillation as the model learns sample the mode of the distributions. However, it significantly reduces inference speed by nearly 90\%, proving the effectiveness and efficiency of our distillation design. Lastly, removing either the multimodal waveform discriminator or multimodal prosody discriminator significantly decreases perceived naturalness and similarity, with the multimodal prosody discriminator having a more substantial impact. This is because the acoustic synthesizer still benefits from MPD and MRD during training, while the prosody autoencoder only relies on $\ell_1$ loss without adversarial training. However, since UT-MOS primarily focuses on the acoustic aspects of speech for naturalness while largely ignores the prosody naturalness, the predicted MOS is unaffected. 

\begin{table}[!h]
\setlength{\tabcolsep}{1mm} 
\small
  \caption{The average percentage of processing time for each module for synthesizing varying lengths of speech. }
  \label{tab:5}
  \centering

  \begin{tabular}{ll}
    \toprule
    Module & \% Time \\ 
    \midrule
    Prompt-Text Encoder & 6.1\% \\
    Distilled Style Diffusion Model & 12.3 \% \\
    Prosody Decoder & 9.9 \% \\
    Acoustic Decoder & 71.7 \% \\
    \bottomrule
  \end{tabular}
\end{table} 

\begin{table*}[!th]
\small
\caption{Comparison of Pearson correlation coefficients of acoustic features associated with emotions between prompt and synthesized speech.}
\vspace{5 pt}

\label{tab:3}

\centering
\begin{tabular}{lccccccccc}
\toprule
Model & \makecell{Pitch \\mean} & \makecell{Pitch \\standard \\deviation} & \makecell{Energy \\mean} & \makecell{Energy \\standard \\deviation} & \makecell{Harmonics-\\to-noise \\ratio}  & \makecell{Speaking \\ rate}
& Jitter & Shimmer   \\
\midrule
Ground Truth    & 0.86 & 0.35 & 0.56 & 0.63 & 0.68 
&0.38 &0.36  & 0.62\\ 
\midrule
NaturalSpeech 2   & 0.88 & 0.41 & 0.81 & 0.40 & 0.83 
&0.03 &0.57  & 0.74\\ 
FlashSpeech & 0.89 & 0.02 & {0.83} & 0.23 & 0.67 
&0.03 &0.64  & 0.49\\ 
VoiceCraft & \textbf{0.96} & \textbf{0.69} & 0.63 & 0.33 & 0.74 
&0.10 &0.72  & 0.76\\ 
XTTSv2 & 0.93 &  0.39 & 0.70 & 0.26 & 0.73 
& 0.45 & 0.84  & 0.63\\ 
\midrule
StyleTTS-ZS (LT)   & 0.94 & 0.53 & 0.74 & 0.32 & \textbf{0.84}
&  0.26 &  0.76  & 0.77 \\ 

StyleTTS-ZS (LL)   & \textbf{0.96} & 0.53 & \textbf{0.88} & \textbf{0.70} & 0.81
&  \textbf{0.46} &  \textbf{0.88}  & \textbf{0.81} \\ 

\bottomrule
\end{tabular}
\end{table*}

\subsection{Other Applications} 

\noindent \textbf{Speech Editing.} By training our prosody decoder with masked prosody, we enable speech editing. To edit a specific part of the speech, we mask the prosody of that segment and decode the prosody conditioned on the new text and input speech as a prompt. This approach retains the original prosody as much as possible while generating new prosody for the edited segment. Re-synthesizing the speech using the edited prosody and the input speech prompt allows for seamless speech editing.

\noindent \textbf{Zero-Shot Voice Conversion. } Our model also supports zero-shot voice conversion conditioned on the text of the input speech, which can be accurately obtained through modern ASR models. We extract the duration, energy, and pitch from the input speech, and the energy and pitch from the prompt. By computing the median of both input and prompt pitch and energy and re-normalizing the input pitch and energy to match the prompt's median values, we ensure consistency. Using the ground truth duration, rescaled pitch, and energy, we can re-synthesize the speech with the prompt speech, achieving effective zero-shot voice conversion. We provide samples on our audio demo page. 

\subsection{Processing Time Analysis}
\label{app:a5}
We analyzed the processing time of each module using the same 40 samples as those used for computing the real-time factor (RTF). The percentage of time taken by each module is shown in Table \ref{tab:5}. The results indicate that the acoustic decoder is the most time-consuming component, suggesting a potential area for improvement. Future work could focus on reducing the acoustic synthesis time by employing ultra-fast acoustic decoders, such as those utilizing inverse short-time Fourier transform (iSTFT), as demonstrated in Voco \cite{siuzdak2023vocos} and HiFT-NET \cite{li2023hiftnet}.

\section{Conclusions and Limitations}
\label{conclusion}

In this work, we propose StyleTTS-ZS, a highly efficient and high-quality zero-shot TTS system that achieves performance on par with previous state-of-the-art (SOTA) models while being 10-20 times faster. Our proposed StyleTTS-ZS model has demonstrated significant advancements in zero-shot TTS synthesis by leveraging novel distilled time-varying style diffusion to efficiently capture diverse speaker identities and prosodies. The results also indicate strong scalability and performance improvements over the LibriTTS model when trained on large-scale datasets such as LibriLight. This model paves the way for various applications, including enhanced real-time human-computer interactions using speech synthesis technology, especially with the integration of large language models that have seen rapid development in recent years \cite{chang2024survey}. For example, StyleTalker \cite{li2024styletalker} has demonstrated an end-to-end integration of style-based TTS models for spoken dialog generation, where StyleTTS-ZS can be a plug-in for the TTS module. This advancement enables the development of customized real-time assistants and virtual companions. Additionally, the reduced computational load required for high-quality speech synthesis makes it ideal for applications such as audiobook narration and other long-form content, improving accessibility tools for individuals with speech impairments by providing more natural and personalized speech outputs. Furthermore, it allows for personalized and dynamic content creation in media, entertainment, and virtual environments.

However, the capabilities of our model also bring potential negative impacts. The ability to generate high-quality, recognizable speech could be misused for voice spoofing, posing risks to personal and financial security. There is also the potential for creating convincing deepfake audio, which can be used maliciously in misinformation campaigns, fraud, or defamation. To mitigate these risks, we recommend controlled access to the model, with strict licensing agreements that require users to obtain consent from individuals whose voices are being cloned. Establishing ethical guidelines and usage policies is essential to prevent misuse. Additionally, encouraging collaboration in the research community to develop and improve deepfake detection technologies and incorporating mechanisms to watermark or trace synthetic audio to ensure accountability and traceability are necessary steps.

Although our model outperforms previous state-of-the-art models, it still has not achieved human-level performance for zero-shot TTS with unseen speakers, as models trained on small-scale datasets for seen speakers have achieved  \cite{tan2024naturalspeech, li2024styletts}. Moreover, our model prioritizes speed over quality, meaning the acoustic synthesizer does not benefit from the latest generative modeling developments with iterative refinements that could further enhance speech quality and fidelity to the prompt speaker. Additionally, the models were trained on English audiobook reading datasets rather than audio from more diverse, real-world environments and other languages. Future research should explore higher-quality generation methods to improve quality without compromising speed, expand training to include more diverse datasets and languages to enhance generalizability, and further refine prosody modeling to achieve even more natural and expressive speech synthesis. Overall, our findings indicate that StyleTTS-ZS is a robust and efficient solution for zero-shot TTS synthesis, with promising potential for applications on large-scale real-world data and future research directions.

\section{Acknowledgments}
We thank Gavin Mischler and Ashley Thomas for assessing the quality of synthesized samples and providing feedback on the quality of models during the development stage of this work. We also thank Kundan Kumar for his feedback on the manuscript. This work was done during the internship at Descript Inc. (Y.A. Li) and was funded by the National Institute of Health (NIHNIDCD) and a grant from Marie-Josee and Henry R. Kravis.

\bibliography{mybib}
\bibliographystyle{IEEEbib}

\clearpage


\begin{appendices}

\section{Additional Results}

\subsection{Prosody Autoencoder Bottleneck}
\label{app:a2}

\begin{figure}[!h]
    \centering
      \subfigure[Effects of codebook size with $K = 50$. ]{\includegraphics[width=0.45\textwidth]{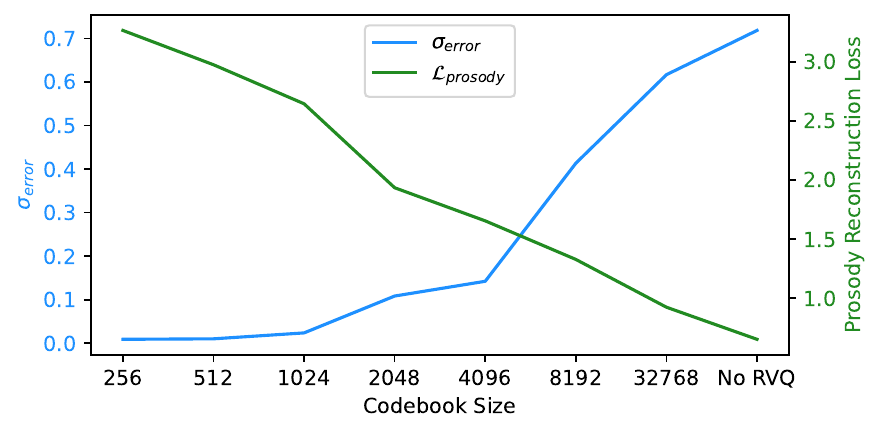}}
      \subfigure[Effects of $K$ with codebooks size of 1024. ]{\includegraphics[width=0.45\textwidth]{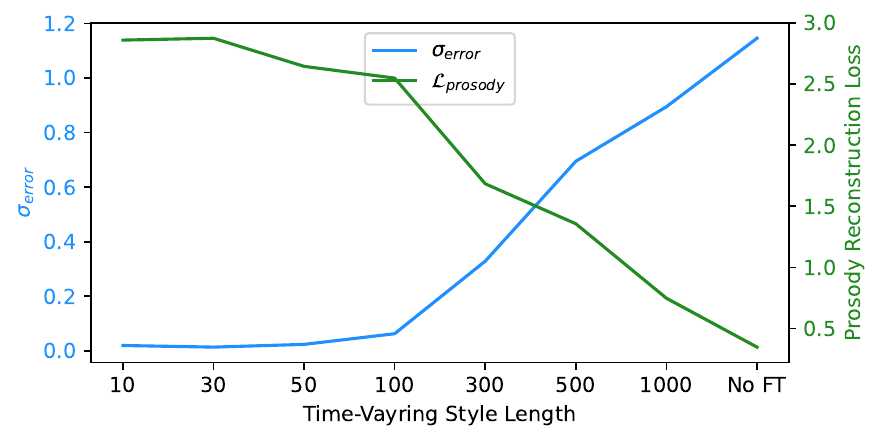}}

    \caption{Effects of bottleneck complexity of prosody autoencoder and diffusion denoiser's performance. (a) Fixed time-varying style length $K = 50$ and varying codebook size. (b) Fixed codebook size of 1024 and varying style length $K$. }
    \label{fig:6}
\end{figure}

We examine the trade-off between the reconstruction error of the prosody autoencoder and the diffusion modeling difficulty by varying the codebook size with the length  $K =50$ of time-varying style and varying length with a fixed codebook size of 1024. We measure the reconstruction error by the aggregated prosody loss $\mathcal{L}_{prosody} = \mathcal{L}_{dur} + \mathcal{L}_{f0} + \mathcal{L}_{n}$ as defined in \ref{app:D2}. We use the normalized standard deviation error as a measure for diffusion modeling complexity: 

\begin{equation}
    \sigma_{\text{error}} = \frac{\norm{ \sigma_{\text{data}} - \sigma_{\text{diff}} }_1}{\sigma_{\text{data}}},
\end{equation}
where $\sigma_{\text{data}}$ is the standard deviation of the data distribution  $\mathcal{D}$ and $\sigma_{\text{diff}}$ is that of diffusion samples after re-normalization. When the diffusion model converges but the error in standard deviation is large, it indicates that the diffusion model is not powerful enough to sample from $\mathcal{D}$ starting from a unit standard deviation distribution $\mathcal{N}(0, 1)$, an observation made in \cite{karras2022elucidating}. 

As shown in Figure \ref{fig:6} (a), higher codebook size results in lower reconstruction error but higher $\sigma_\text{error}$, with reconstruction error being the lowest while $\sigma_\text{erorr}$ being the highest when no RVQ is used. This finding justifies the importance of using RVQ in our prosody autoencoder design. Similarly, with a fixed codebook size, the higher the length of the time-varying style, the lower the reconstruction error but the higher the $\sigma_\text{error}$. When varying length style vector is used as in the ``No FT" case, the prosody encoder is the same as just applying RVQ to the input prosody, and thus, the diffusion complexity is proportional to the input size. This finding motivates us to apply adversarial training with multimodal discriminators for our prosody autoencoder training in order to shift the burden from the diffusion model to the prosody decoder for efficient sample generation. 

\subsection{Diffusion Distillation}
\label{app:a3}

We compare our simulation-based distillation with perceptual loss to simulation-free bootstrapping methods such as consistency distillation \cite{song2023consistency} and adversarial diffusion distillation \cite{sauer2023adversarial} with varying sample sizes and whether the student network is initialized with a pre-trained model that predicts the ground truth time-varying style conditioned on the input text $\bm{t}$ and prompt $\bm{x}'$. We use the $\ell_1$ distance of the decoded duration ($\mathcal{L}_{\text{dur}}$), pitch ($\mathcal{L}_{f0}$) and energy ($\mathcal{L}_{n}$) between teacher and student models from the same input noise conditioned on unseen text and speaker prompts as the metric to measure the distillation performance. We tested the model performance using the guidance $\omega = 5$ as during our inference process. 

\begin{table*}[!h]
\small
\caption{Comparison of $\mathcal{L}_{\text{dur}}$, $\mathcal{L}_{f0}$ and $\mathcal{L}_{n}$ with various distillation methods and sample size. The method used in our experiments is highlighted in italics. }
\label{tab:13}
\centering
\begin{tabular}{lcccc}
\toprule
Method & Sample size & $\mathcal{L}_{\text{dur}}$ & $\mathcal{L}_{f0}$ & $\mathcal{L}_{n}$ \\
\midrule
Consistency distillation \cite{song2023consistency} & --- & 0.34 & 0.83 & 0.36 \\
Adversarial diffusion distillation \cite{sauer2023adversarial} & --- & 0.29 & 0.77 & 0.24 \\
\midrule
Simulation-based (w/ pre-trained initialization) & 1k & 0.37 & 1.07 & 0.34 \\
Simulation-based (w/ pre-trained initialization) & 5k & 0.29 & 0.93 & 0.29 \\
\textit{Simulation-based (w/ pre-trained initialization)} & \textit{10k} & \textit{0.18} & \textit{0.64} & \textit{0.18} \\
Simulation-based (w/ pre-trained initialization) & 30k & 0.16 & 0.52 & 0.13 \\
Simulation-based (w/ pre-trained initialization) & 50k & 0.13 & \textbf{0.43} & 0.09 \\
Simulation-based (w/ pre-trained initialization) & 100k & \textbf{0.11} & 0.44 & \textbf{0.07} \\
\midrule
Simulation-based (w/o pre-trained initialization) & 1k & 0.65 & 1.74 & 0.59 \\
Simulation-based (w/o pre-trained initialization) & 5k & 0.58 & 1.63 & 0.52\\
Simulation-based (w/o pre-trained initialization) & 10k & 0.53 & 1.35 & 0.44 \\
Simulation-based (w/o pre-trained initialization) & 30k & 0.44 &  1.02 & 0.29 \\
Simulation-based (w/o pre-trained initialization) & 50k & 0.38 & 0.81 & 0.24 \\
Simulation-based (w/o pre-trained initialization) & 100k & 0.24 & 0.67 & 0.18 \\
\bottomrule
\end{tabular}
\end{table*}

For the consistency distillation (CD) baseline, we follow Algorithm 1 in \cite{luo2023latent} for guided distillation. We use the DDIM solver with noise schedule specified in section \ref{sec:3.3}, $\ell_1$ norm for distance, EMA rate $\mu = 0.999943$ as in \cite{song2023consistency} and guidance scale range $\omega_\text{min} = 1, \omega_\text{max} = 15$ as in equation \ref{eq:5}. For the adversarial diffusion distillation (ADD) baseline, we use the architecture of multimodal prosody discriminator as the discriminator that takes the denoised time-varying style as input and text embeddings $\bm{t}$ and prompt $\bm{x}'$ as conditions. We set $\tau_n = 200$ as our simulation-based method uses 100 steps and the student step $N = 4$ as in \cite{sauer2023adversarial}. We initialized the student network with the teacher network for both CD and ADD baselines as in \cite{sauer2023adversarial}. For the case of the simulation-based approach without pre-trained initialization, we also initialized our student network with the teacher network for fair comparison. We trained our model for 100k steps with a batch size of 32. 

As shown in Table \ref{tab:13}, our simulation-based approach with only 10k samples achieves better results than both CD and ADD baselines and much lower perceptual discrepancies compared to student models without the pre-trained initialization. This is likely because the classifier-free guidance is not easily approximated with bootstrapping-based methods, as their effects depend on fine-grained trajectories with small step sizes. These bootstrapping-based methods are useful for distilling large-scale diffusion models that are expensive to run sampling processes to get samples directly. However, since our latent is a fixed-size $50 \times 512$ vector, it is straightforward to run simulations and obtain enough samples to cover the latent space, particularly with our novel initialization approach with pre-trained networks that directly predict the ground truth latent. 

\section{Evaluation Details}
\label{app:B}
\subsection{Baseline Models and Samples}
\label{app:B1}

In this section, we provide brief introductions to each baseline model and our methods to obtain samples to conduct our evaluations. 

\begin{itemize}
    \item \textbf{Vall-E}: Vall-E \cite{wang2023neural} is a previous state-of-the-art (SOTA) model trained on the LibriLight dataset that consists of 60k hours of speech data. This is also the first large-scale zero-shot TTS model using language modeling techniques in large language model (LLM) research. Since this model is not publicly available, we obtained 40 samples from the authors along with the text transcription and 3-second prompt speech to synthesize the corresponding speech of our models for comparison. These samples were also used to compute the speaker embedding similarity (SIM) and word error rate (WER).

    \item \textbf{NaturalSpeech 2}: NaturalSpeech 2 \cite{shen2024naturalspeech} is a previous SOTA zero-shot TTS model trained on Multilingual LibriSpeech (MLS) dataset \cite{pratap2020mls} with 44k hours of audio that shows strong performance on zero-shot speech synthesis with high quality and fidelity. Since this model is not publicly available, we obtained 40 samples from the authors along with the text transcription and 3-second prompt speech to synthesize the corresponding speech of our models for comparison. These 40 samples have the same texts and speech prompts as provided by Vall-E authors. These samples were also used to compute the speaker embedding similarity (SIM) and word error rate (WER).

    \item \textbf{FlashSpeech}: FlashSpeech \cite{ye2024flashspeech} is the current SOTA TTS model trained on the Multilingual LibriSpeech (MLS) dataset with significantly faster inference speed compared to most other large-scale high-quality zero-shot TTS models. Since this model is not publicly available, we obtained 40 samples from the authors along with the text transcription and 3-second prompt speech to synthesize the corresponding speech of our models for comparison. These 40 samples have the same texts and speech prompts as provided by Vall-E and NaturalSpeech 2 authors. These samples were also used to compute the speaker embedding similarity (SIM) and word error rate (WER).

    \item \textbf{NaturalSpeech 3}: NaturalSpeec 3 \cite{ju2024naturalspeech} is the current state-of-the-art (SOTA) model in zero-shot speech synthesis trained on LibriLight with factorized codec and diffusion models. It has achieved close-to-human performance in terms of prompt speaker similarity. Since this model is not publicly available, we obtained 40 samples from the authors along with the text transcription and 3-second prompt speech to synthesize the corresponding speech of our models for comparison. These 40 samples have the same texts and speech prompts as provided by Vall-E and NaturalSpeech 2 authors. These samples were also used to compute the speaker embedding similarity (SIM) and word error rate (WER). 

    \item \textbf{VoiceCraft}: VoiceCraft \cite{peng2024voicecraft} is a strong autoregressive baseline model for zero-shot speech synthesis trained on the GigaSpeech dataset that exhibits high performance in terms of speaker similarity for prompt speech in the wild. This model is publicly available at \url{https://github.com/jasonppy/VoiceCraft} and we synthesized 40 samples using the same text and 3-second speech prompts as provided by the authors of \cite{wang2023neural, shen2024naturalspeech, ye2024flashspeech} for fair comparison. 

    \item \textbf{XTTSv2}: XTTSv2 \cite{casanova2022yourtts} is a strong baseline model for zero-shot speech synthesis trained on the various public datasets. This model is publicly available at \url{https://huggingface.co/coqui/XTTS-v2} and we synthesized 40 samples using the same text and 3-second speech prompts as provided by the authors of \cite{wang2023neural, shen2024naturalspeech, ye2024flashspeech} for fair comparison. 

    \item \textbf{StyleTTS 2}: StyleTTS 2 \cite{li2024styletts} is a SOTA model for single-speaker synthesis with fast inference speed and the capability to perform high-quality zero-shot speech synthesis. This model is publicly available at \url{https://github.com/yl4579/StyleTTS2} and we synthesized 40 samples using the same text and 3-second speech prompts as provided by the authors of \cite{wang2023neural, shen2024naturalspeech, ye2024flashspeech} for fair comparison with the LibriTTS checkpoint at \url{https://huggingface.co/yl4579/StyleTTS2-LibriTTS/tree/main}. 

    \item \textbf{HierSpeech++}: HierSpeech++ \cite{lee2023hierspeech++} is a strong baseline model that exhibits high speaker similarity with limited training sets. This model is publicly available at \url{https://github.com/sh-lee-prml/HierSpeechpp}. We synthesized 40 samples using the same text and 3-second speech prompts as provided by the authors of \cite{wang2023neural, shen2024naturalspeech, ye2024flashspeech} for a fair comparison with the \textit{LibriTTS-train-960} checkpoint.

 \end{itemize}

\subsection{Subjective Evaluations}

To ensure high-quality evaluation from MTurk, we followed \cite{li2024styletts} by enabling the following filters on MTurk:
\begin{itemize}
    \item HIT Approval Rate (\%) for all Requesters' HITS: \verb|greater than 95|.
    \item Location: \verb|is UNITED STATES (US)|.
    \item Number of HITs Approved: \verb|greater than 50|.
\end{itemize}

We provided the following instructions for rating the naturalness and similarity of our MOS experiments following \cite{li2024styletts}:
\begin{itemize}
    \item Naturalness: \begin{verbatim}Some of them may be synthesized while
others may be spoken by an American 
audiobook narrator.

Rate how natural each audio clip 
sounds on a scale of 1 to 5 with 1 
indicating completely unnatural 
speech (bad) and 5 completely natural
speech (excellent). 

Here, naturalness includes whether you 
feel the speech is spoken by a native 
American English speaker from a human 
source.\end{verbatim}
    
    \item Similarity: \begin{verbatim}Rate whether the two audio clips could 
have been produced by the same speaker
or not on a scale of 1 to 5 with 1 
indicating completely different 
speakers and 5 indicating exactly the
same speaker.

Some samples may sound somewhat 
degraded/distorted; for this question,
please try to listen beyond the 
distortion of the speech and 
concentraten on identifying the voice 
(including the person's accent and 
speaking habits, if possible).\end{verbatim}
\end{itemize}
An example survey used for our MOS evaluation can be found at \url{https://survey.alchemer.com/s3/7858362/styletts-new-MOS-521}.

For CMOS experiments, since all the models have high-quality synthesis results, we removed the instruction regarding audio distortion and used the following instructions instead:
\begin{itemize}
\item Naturalness: \begin{verbatim}Some of them may be synthesized while 
others may be spoken by an American
audiobook narrator.

Rate how natural is B compared to A on 
a scale of -6 to 6, with 6 indicating 
that B is much better than A. 

Here, naturalness includes whether 
you feel the speech is spoken by a 
native American English speaker from 
a human source.\end{verbatim}
    
\item Similarity: \begin{verbatim}Rate how similar is the speaker in B 
to the reference voice, compared to A. 
Here, "similar" means that you feel 
that the recording and the reference 
voice are produced by the same speaker.
\end{verbatim}
\end{itemize}

An example survey used for our CMOS evaluation is available at \url{https://survey.alchemer.com/s3/7854889/CMOS-stylettsz-flashspeech-librispeech-0519}. 

We ensured the survey quality by applying additional attention checking following \cite{li2024styletts}. In the MOS assessment, we utilized the average score given by a participant to ground truth audios, unbeknownst to the participants, to ascertain their attentiveness. We excluded ratings from those whose average score for the ground truth did not rank in the top three among all five models. In the CMOS evaluation, we checked the consistency of the rater's scores: if the score's sign (indicating whether A was better or worse than B) differed for over half the sample set, the rater was disqualified. Four raters were eliminated through this process in all of our experiments.

For the CMOS experiments with \cite{wang2023neural, shen2024naturalspeech, ye2024flashspeech, peng2024voicecraft} and ground truth, since we have 40 samples, we divided it into two batches with 20 samples each. We recruited 10 raters for the first batch, which consisted of 20 pairs of samples, and obtained 200 ratings from this batch. We then launched a second batch with another 20 pairs of samples and obtained another 200 ratings. For other models, since we did not have enough samples, we recruited more raters until we had 400 ratings, excluding disqualified raters. For example, for the experiments with VoiceBox, as we have only 15 samples, we recruited 28 raters after one rater was disqualified.

\label{app:B2}

\section{Training Objectives}
\label{app:D}

In this section, we provide detailed training objectives for our acoustic synthesizer and prosody autoencoder, as the training objective for style diffusion is provided in Section \ref{sec:3.3}. 

\subsection{Acoustic Synthesizer Training}
In this section, we follow the notation of \cite{li2024styletts} where the decoder $G(\cdot)$ takes upsampled text embeddings $\bm{h}_{\text{text}} \cdot \bm{a}$, pitch $p_{\bm{x}}$, energy $n_{\bm{x}}$, and global style $\bm{s}$ instead of five inputs as in Section \ref{sec:3.1} for consistency with \cite{li2024styletts}. In section \ref{sec:3.1}, we use duration $d_{\bm{x}}$ as a compact representation for the phoneme-speech alignment $\bm{a}$, which is what is being used in the actual implementation. $\bm{a}$ can be obtained by repeating the value 1 for $d_i$ times at $\ell_{i-1}$, where $\ell_{i}$ is the end position of the $i^\text{th}$ phoneme $\bm{t}_i$ calculated by summing $d_k$ for $k \in \{1, \ldots, i\}$, and $d_i$ is the duration of $\bm{t}_i$.

\textbf{Mel-spectrogram reconstruction.} The decoder is trained on waveform $\bm{y}$, its corresponding mel-spectrogram $\bm{x}$, a speech prompt mel-spectrogram $\bm{x}'$ which is a small chunk of $\bm{x}$, and the text $\bm{t}$, using $L_1$ reconstruction loss as 

\begin{equation} \label{eq1}
\mathcal{L}_\text{mel} = \mathbb{E}_{\bm{x}, \bm{t}}\left[{\norm{\bm{x} - M\parn{G\parn{\bm{h}_\text{text} \cdot \bm{a}_\text{algn}, \bm{s}, p_{\bm{x}}, n_{\bm{x}}}}}_1}\right].
\end{equation}

Here, $\bm{h}_\text{text}, \bm{s} = T(\bm{t}, \bm{x}')$ is the encoded phoneme representation aligned with the prompt $\bm{x}'$ and global style of $\bm{x}'$, and the attention alignment is denoted by $\bm{a}_\text{algn} = A(\bm{x}, \bm{t})$. $p_{\bm{x}}$ is the pitch F0 and $n_{\bm{x}}$ indicates energy of $\bm{x}$, which are extracted by a pitch extractor $p_{\bm{x}} = F(\bm{x})$, and $M(\cdot)$ represents mel-spectrogram transformation. Following \cite{li2022styletts}, half of the time, raw attention output from $A$ is used as alignment, allowing backpropagation through the text aligner. For another 50\% of the time, a monotonic version of $\bm{a}_\text{algn}$ is utilized via dynamic programming algorithms (see Appendix A in \cite{li2022styletts}).

\textbf{TMA objectives. } We follow \cite{li2022styletts} and use the original sequence-to-sequence ASR loss function $\mathcal{L}_\text{s2s}$ to fine-tune the pre-trained text aligner, preserving the attention alignment during end-to-end training:

\begin{equation} 
\label{eq3}
\mathcal{L}_\text{s2s} = \mathbb{E}_{\bm{x}, \bm{t}}\left[\sum\limits_{i=1}^N{\textbf{CE}(\bm{t}_i, \hat{\bm{t}}_i)}\right] ,
\end{equation}

\noindent where $N$ is the number of phonemes in $\bm{t}$, $\bm{t}_i$ is the $i$-th phoneme token of $\bm{t}$, $\hat{\bm{t}}_i$ is the $i$-th predicted phoneme token, and $\textbf{CE}(\cdot)$ denotes the cross-entropy loss function.

Additionally, we apply the monotonic loss $\mathcal{L}_\text{mono}$ to ensure that soft attention approximates its non-differentiable monotonic version:

\begin{equation} \label{eq4}
\mathcal{L}_\text{mono} = \mathbb{E}_{\bm{x}, \bm{t}}\left[\norm{\bm{a}_\text{algn} - \bm{a}_\text{hard}}_1\right] ,
\end{equation}

where $\bm{a}_\text{hard}$ is the monotonic version of $\bm{a}_\text{algn}$ obtained through dynamic programming algorithms (see Appendix A in \cite{li2022styletts} for more details).

\textbf{Adversarial objectives. } Two adversarial loss functions, originally used in HifiGAN \cite{kong2020hifi}, are employed to enhance the sound quality of the reconstructed waveforms: the LSGAN loss function $\mathcal{L}_\text{adv}$ for adversarial training and the feature-matching loss $\mathcal{L}_\text{fm}$. 

\begin{equation}
\begin{split}
      & \mathcal{L}_\text{adv} (G;D) = \\
      & \mathbb{E}_{ \bm{t}, \bm{x}}\left[\parn{D\parn{\parn{{G\parn{\bm{h}_\text{text} \cdot \bm{a}_\text{algn}, \bm{s}, p_{\bm{x}}, n_{\bm{x}}}}}; \mathcal{C}} - 1}^2\right], \\
\end{split}
\end{equation}

\begin{equation}
\begin{aligned}    
  & \mathcal{L}_\text{adv}(D;G) =\text{ } \\
  &\mathbb{E}_{\bm{t}, \bm{x}}\left[\parn{D\parn{\parn{G\parn{\bm{h}_\text{text} \cdot \bm{a}_\text{algn}, \bm{s}, p_{\bm{x}}, n_{\bm{x}}}}} ; \mathcal{C}}^2\right] + \\ &\mathbb{E}_{\bm{y}}\left[
   \parn{D(\bm{y}; \mathcal{C}) - 1}^2\right],
\end{aligned}
  \label{eq1}
\end{equation}

\begin{equation} \label{eq6}
\begin{aligned}  
\mathcal{L}_\text{fm} = \mathbb{E}_{\bm{y}, \bm{t}, \bm{x}}\left[\sum\limits_{l=1}^{\Lambda} \frac{1}{N_l}\norm{D^l(\bm{y}; \mathcal{C}) - D^l\parn{\bm{\hat{y}}; \mathcal{C}}}_1\right],
\end{aligned}
\end{equation}

\noindent where $\bm{\hat{y}} = G\parn{\bm{h}_\text{text} \cdot \bm{a}_\text{algn}, \bm{s}, p_{\bm{x}}, n_{\bm{x}}}$ is the generated waveform and $D$ represents both MPD, MRD, and multimodal waveform discriminator (MMWD). $\mathcal{C}$ denotes the conditional input to MMWD (see Section \ref{sec:3.4}) and is $\emptyset$ for MPD and MRD. $\Lambda$ is the total number of layers in $D$, and $D^l$ denotes the output feature map of $l$-th layer with $N_l$ number of features.   

\textbf{Speaker Embedding Feature Matching Loss. } We compute the intermediate features of a ResNet-based speaker embedding model \cite{wang2023wespeaker} $V$ for the following loss: 

\begin{equation} \label{eq:sv}
\begin{aligned}  
\mathcal{L}_\text{fm} = \mathbb{E}_{\bm{y}, \bm{t}, \bm{x}}\left[\sum\limits_{l=1}^{\Lambda} \frac{1}{N_l}\norm{V^l(\bm{y}) - V^l\parn{\bm{\hat{y}}}}_1\right],
\end{aligned}
\end{equation}

\noindent $\bm{\hat{y}} = G\parn{\bm{h}_\text{text} \cdot \bm{a}_\text{algn}, \bm{s}, p_{\bm{x}}, n_{\bm{x}}}$ is the generated waveform, $\Lambda$ is the total number of layers in $V$, $V^{l}$ denotes the output feature map of $l$-th layer with $N_l$ number of features.  

\textbf{Acoustic synthesizer full objectives. } Our full objective functions in acoustic synthesizer training can be summarized as follows with hyperparameters $\lambda_\text{s2s}$ and $\lambda_\text{mono}$:

\begin{equation} \label{eq7}
\begin{aligned}  
\min_{G, A, T, F} \text{    }&
 \mathcal{L}_\text{mel} + \lambda_\text{s2s}\mathcal{L}_\text{s2s} +  \lambda_\text{mono}\mathcal{L}_\text{mono}  \\ & 
+ \mathcal{L}_\text{adv}(G;D) + \mathcal{L}_\text{fm}
\end{aligned}
\end{equation}
\begin{equation} \label{eq7}
\begin{aligned}  
\min_{D} \text{    } &\mathcal{L}_\text{adv}(D;G) 
\end{aligned}
\end{equation}

Following \cite{li2022styletts}, we set $\lambda_\text{s2s} = 0.2$ and $\lambda_\text{mono} = 5$.

\subsection{Prosody Autoencoder Training}
\label{app:D2}
\textbf{Duration reconstruction. } We employ the $L$-1 loss to reconstruct the duration:

\begin{equation} \label{eq8}
\mathcal{L}_\text{dur} = \mathbb{E}_{\bm{x}}\left[\norm{{d}_{\bm{x}} - \hat{d}_{\bm{x}}}_1\right],
\end{equation}

\noindent where $\hat{d}_{\bm{x}} = \text{PD}_d(\text{PAE}(d_{\bm{x}}, p_{\bm{x}}, n_{\bm{x}}), \bm{t})$ is the decoded duration  conditioned on $\bm{t}$ from the duration decoder $\text{PD}_d(\cdot)$ after being encoded by the prosody encoder $\text{PAE}(\cdot)$. 

\textbf{Prosody reconstruction. }  We use $\mathcal{L}_{f_0}$ and $\mathcal{L}_n$, which are F0 and energy reconstruction loss, respectively: 

\begin{equation} \label{eq9}
\mathcal{L}_{f0} = \mathbb{E}_{{\bm{x}}}\left[\norm{{p}_{\bm{x}} - \hat{p}_{\bm{x}}}_1\right]
\end{equation}

\begin{equation} \label{eq10}
\mathcal{L}_{n} = \mathbb{E}_{{\bm{x}}}\left[\norm{{n}_{\bm{x}} - \hat{n}_{\bm{x}}}_1\right]
\end{equation}

\noindent where $\hat{p}_{\bm{x}}, \hat{n}_{\bm{x}} = \text{PD}_p(\text{PAE}(d_{\bm{x}}, p_{\bm{x}}, n_{\bm{x}}), \bm{t})$ are the decoded pitch and energy of $\bm{x}$  conditioned on $\bm{t}$ from the pitch and energy decoder $\text{PD}_p(\cdot)$. 

\textbf{Adversarial objectives. } We employed similar adversarial objectives as during the acoustic synthesizer training: 

\begin{equation}
      \mathcal{L}_\text{adv}(G;D) =\text{ } \mathbb{E}_{ \bm{t}, \bm{x}}\left[\parn{D\parn{\parn{{G\parn{\mathcal{P}}}}; \mathcal{C}} - 1}^2\right], \\
\end{equation}

\begin{equation}
\begin{aligned}    
  \mathcal{L}_\text{adv}(D;G) =\text{ }
  &\mathbb{E}_{\bm{t}, \bm{x}}\left[\parn{D\parn{\parn{G\parn{\mathcal{P}}}} ; \mathcal{C}}^2\right] \\ & +   \mathbb{E}_{\bm{y}}\left[
   \parn{D(\mathcal{P}; \mathcal{C}) - 1}^2\right],
\end{aligned}
  \label{eq1}
\end{equation}

\begin{equation} \label{eq6}
\mathcal{L}_\text{fm} = \mathbb{E}_{\bm{y}, \bm{t}, \bm{x}}\left[\sum\limits_{l=1}^{\Lambda} \frac{1}{N_l}\norm{D^l(\mathcal{P}; \mathcal{C}) - D^l\parn{G\parn{\mathcal{P}}; \mathcal{C}}}_1\right],
\end{equation}

\noindent where $D$ represents the multimodal prosody discriminator (MMPD) for duration and pitch and energy and $G$ represents the combined prosody encoder and decoder $\text{PD}(\text{PAE}(\cdot), \bm{t})$. $\mathcal{C}$ denotes the conditional input to MMPD and $\mathcal{P}$ denotes either duration, pitch or energy. $\Lambda$ is the total number of layers in $D$, and $D^l$ denotes the output feature map of $l$-th layer with $N_l$ number of features.

\textbf{Prosody autoencoder training full objectives. } Our full objective functions in joint training can be summarized as follows with hyperparameters $\lambda_\text{dur}, \lambda_{f0}$, and $\lambda_{n}$:

\begin{equation} \label{eq7}
\begin{aligned}  
\min_{G, D} \text{    }&
 \mathcal{L}_\text{dur} + \lambda_\text{f0}\mathcal{L}_\text{f0} +  \lambda_\text{n}\mathcal{L}_\text{n}  \\ & 
+ \mathcal{L}_\text{adv}(G;D) + \mathcal{L}_\text{fm}
\end{aligned}
\end{equation}
\begin{equation} \label{eq7}
\begin{aligned}  
\min_{D} \text{    } &\mathcal{L}_\text{adv}(D;G) 
\end{aligned}
\end{equation}
\noindent where $G$ represents both prosody encoder and decoder. Following \cite{li2024styletts}, we set $\lambda_{f0} = 0.1$ and $\lambda_{n} = 1$. 

\section{Model Architectures}
\label{app:C}
This section provides a detailed outline of StyleTTS-ZS architecture. We keep the same architecture for the waveform decoder, duration extractor (or text aligner in \cite{li2024styletts}), and pitch extractor as in \cite{li2024styletts}. The prosodic text encoder is a pre-trained PL-BERT \cite{li2023phoneme} available at \url{https://github.com/yl4579/PL-BERT}. Additionally, we adopt the same acoustic discriminators as in \cite{kumar2024high}. This section focuses primarily on our new proposed components as outlined in Figure \ref{fig:2}. 

\begin{table*}[!hbp]
\caption{Prompt-text encoder architecture. $N$ represents the input phoneme length of the mel-spectrogram and $T$ represents the prompt length, $\bm{t}$ represents the input phonemes with shape $1 \times N$, $\bm{x}'$ is the speech prompt mel-spectrogram with shape $80 \times T$. }
\centering
\begin{tabular}{cccc}
\hline
Submodule                          & Input                    & Layer       & Output Shape      \\ 
\hline
\multirow{3}{*}{Embedding}     & $\bm{t}$                 & Phoneme  Embedding     &  512 $\times N$               \\
& $\bm{x}'$                & Linear $80 \times 512$         & 512 $\times T$  \\
& $-$                  & Concat       &  $512 \times (N + T) $ \\
\hline
Conformer Block $\parn{\times 1}$
& $-$     & \makecell{\# of head: 8, \\
                    head features: 64, \\
                    kernel size: 31, \\
                    feedforward dimension: 1024
}        & 512 $\times (N + T)$               \\
\hline
Conformer Block $\parn{\times 5}$
& $-$     & \makecell{\# of head: 8, \\
                    head features: 64, \\
                    kernel size: 15, \\
                    feedforward dimension: 1024
}        & 512 $\times (N + T)$               \\

\hline
\multirow{3}{*}{Output}  & $-$  & Split w.r.t. $\bm{t}$ and $\bm{x}'$ &  \makecell{$512 \times N$ \\ $512 \times T$} \\
& $-$  & Adaptive Average Pool w.r.t. $\bm{x}'$ & $512 \times 1$ \\
& $-$  & Linear $512 \times 512$ w.r.t. $\bm{t} $ & $512 \times T$ \\
\hline
\end{tabular}
\end{table*}

\begin{table*}[!hbp]
\caption{Prosody encoder architecture. $T$ represents the length of pitch $p$ and energy $n$ with shape $1 \times T$, while $N$ represents the length of duration $d$ with shape $1 \times N$.  $k_l$ represents position inputs from $\{1, \ldots, l\}$. }
\centering
\begin{tabular}{cccc}
\hline
Submodule                          & Input                    & Layer       & Output Shape      \\ 
\hline
\multirow{3}{*}{Input}     & $d, k_T$                 & Position Upsampling with $d$ and Embedding     &  512 $\times T$               \\
& $p, n$                & Concat         & 514 $\times T$  \\
& $-$                & Linear $514 \times 512$         & 512 $\times T$  \\
\hline
Conformer Block $\parn{\times 1}$
& $-$     & \makecell{\# of heads: 8, \\
                    head features: 64, \\
                    kernel size: 31, \\
                    feedforward dimension: 1024
}        & 512 $\times T$               \\
\hline
\makecell{Conformer Block $\parn{\times 5}$ \\ \\
(output denoted as $\bm{h}_{vl}$)
}
& $-$     & \makecell{\# of heads: 8, \\
                    head features: 64, \\
                    kernel size: 15, \\
                    feedforward dimension: 1024 
}        & 512 $\times T$               \\
\hline
\multirow{3}{*}{Output}  & $k_{50}$ & Embedding &  $512 \times 50$ \\
& $\bm{h}_{vl}$ & 8-Head Cross Attention with 64 head features & $512 \times 50$ \\
\hline
\end{tabular}
\end{table*}

\begin{table*}[!hbp]
\caption{Prosody decoder architecture. $\bm{t}$ represents the input text embeddings from PL-BERT with size $512 \times N$ where $N$ is the text length. $\bm{h}$ represents the time-varying style output from the prosody encoder with size $512 \times 50$. $k = 2$ for pitch and energy decoder and $k = 1$ for duration decoder.}
\centering
\begin{tabular}{cccc}
\hline
Submodule                          & Input                    & Layer       & Output Shape      \\ 
\hline
\multirow{1}{*}{Input}     & $\bm{t}, \bm{h}$                 & Concat     &  512 $\times (N + 50)$               \\
\hline
Conformer Block $\parn{\times 6}$
& $-$    & \makecell{\# of heads: 8, \\
                    head features: 64, \\
                    kernel size: 31, \\
                    feedforward dimension: 1024
}        & 512 $\times (N  + 50)$               \\
\hline
\multirow{3}{*}{Output}  & $-$ & Truncate w.r.t. $\bm{t}$ &  $512 \times N$ \\
& $-$ & Linear $512 \times k$ & $k \times N$ \\
\hline
\end{tabular}
\end{table*}

\begin{table*}[!hbp]
\caption{Style diffusion denoiser architecture. We used the same architecture for the distilled student model. $\bm{t}$ represents the input text embeddings from PL-BERT with size $512 \times N$ where $N$ is the text length and $\bm{x}'$ is the prompt speech with size $80 \times T$. $\bm{\xi}$ represents the noisy input with the  size $512 \times 50$. $\sigma$ represents either the noise level (for the teacher diffusion model) or the guidance scale (for the distilled student model). During pre-training of student model, $\sigma = 0$ and $\bm{\xi} = 0$. }
\centering
\begin{tabular}{cccc}
\hline
Submodule                          & Input                    & Layer       & Output Shape      \\ 
\hline
\multirow{2}{*}{Input}   & $\bm{x}'$                & Linear $80 \times 512$         & 512 $\times N$  \\
 & $\bm{t}, \bm{\xi}$                 & Concat as output $\bm{x}$    &  512 $\times (T + N + 50)$               \\
\hline
\multirow{2}{*}{Embedding}    & $\bm{\sigma}$                 & Sinusoidal  Embedding     &  512 $\times 1$               \\  
& $-$                  & Repeat     &  512 $\times (T + N + 50)$               \\  
& $\bm{x}$                 & Addition     &  512 $\times (T + N + 50)$               \\  

\hline
Conformer Block $\parn{\times 2}$
& $-$     & \makecell{\# of head: 8, \\
                    head features: 64, \\
                    kernel size: 31, \\
                    feedforward dimension: 1024
}        & 512 $\times (N + T + 50)$               \\
\hline
Conformer Block $\parn{\times 10}$
& $-$     & \makecell{\# of head: 8, \\
                    head features: 64, \\
                    kernel size: 15, \\
                    feedforward dimension: 1024
}        & 512 $\times (N + T + 50)$               \\

\hline
Output  & $-$  & Truncate w.r.t. $\bm{\xi}$ &  \makecell{$512 \times 50$} \\
\hline
\end{tabular}
\end{table*}

\begin{table*}[!hbp]
\caption{Multimodal discriminator architecture. $\bm{x}$ represents the decoder output with shape $d \times N$ and $\mathcal{C}$ represents conditions for the discriminator with shape $k \times T$.}
\centering
\begin{tabular}{cccc}
\hline
Submodule                          & Input                    & Layer       & Output Shape      \\ 
\hline
\multirow{3}{*}{Input}     & $\bm{x}$                 & Linear $d \times 512$     &  $512 \times N$               \\
    & $\mathcal{C}$                 & Linear $k \times 512$     &  $512 \times T$               \\
    & $-$  & Concat     &  $512 \times (T + N)$               \\
\hline

Conformer Block $\parn{\times 6}$
& $-$     & \makecell{\# of heads: 8, \\
                    head features: 64, \\
                    kernel size: 15, \\
                    feedforward dimension: 1024
}        & 512 $\times (T  + N)$               \\
\hline
Output   & $-$  & Linear  $1024 \times 1$   &  $1 \times (T + N)$               \\
\hline
\end{tabular}
\end{table*}
\end{appendices}

\end{document}